\documentclass{article}

\usepackage{PRIMEarxiv}

\usepackage[utf8]{inputenc} 
\usepackage[T1]{fontenc}    
\usepackage{hyperref}       
\usepackage{url}            
\usepackage{booktabs}       
\usepackage{amsfonts}       
\usepackage{nicefrac}       
\usepackage{microtype}      
\usepackage{lipsum}
\usepackage{fancyhdr}       
\usepackage{graphicx}       
\graphicspath{{media/}}     
\usepackage{tabularx}
\usepackage{array}
 \usepackage{amsmath} 
\title{DER-Fault: A Simulation-Based Fault Dataset for DER-Integrated Distribution Systems Under Diverse Operating Conditions
}

\author{
  Fathima Razeeya Mohamed Razick, Petr Musilek \\
  Department of Electrical and Computer Engineering \\
  University of Alberta \\
  Edmonton, AB, Canada\\
  \texttt{\{fmohame4, pmusilek\}@ualberta.ca} \\
}

\begin{document}
\maketitle

\begin{abstract}
The growth of distributed energy resource (DER) integration introduces greater variability in distribution system operating conditions, creating a need for comprehensive datasets for developing and evaluating data-driven fault diagnostic methods. This paper presents \textbf{DER-Fault}, a simulation-based dataset generated in OpenDSS using the Iowa 240-bus distribution system with photovoltaic (PV) generation, electric vehicle (EV) charging, and vehicle-to-grid (V2G) operation. To capture annual load variability while maintaining a manageable number of simulation cases, 60 representative operating conditions are selected from 8,760 hourly load profiles, covering peak, off-peak, transition, typical weekday, and typical weekend conditions. Six DER scenarios with varying PV and EV penetration levels and V2G participation are considered. The dataset includes normal operation and ten short-circuit fault classes, with varying fault resistance. Pre- and post-fault voltage and line-current phasor measurements are recorded at selected measurement locations, resulting in 19,800 simulation cases. Technical validation demonstrates broad coverage of annual loading and DER operating conditions, distinct phase-specific responses corresponding to the simulated fault types, and a consistent reduction in voltage-phasor disturbance with increasing fault resistance. The resulting dataset provides a structured resource for developing and evaluating machine-learning and other data-driven methods for fault detection, classification, and localization in DER-integrated distribution systems.
\end{abstract}


\section{Value of the data}
\begin{itemize}
\item The increasing integration of distributed energy resources (DERs) introduces greater variability in distribution-system operating conditions, creating challenges for fault diagnostic approaches. This dataset provides fault measurements under varying loading and DER conditions, supporting the development and evaluation of fault detection methods for DER-integrated distribution networks.
\item The data capture variations in system loading and DER integration, including photovoltaic (PV) generation, electric vehicle (EV) charging, and vehicle-to-grid (V2G) operation, and can therefore support investigations of how changing operating conditions affect fault characteristics and diagnostic performance.
\item The dataset includes multiple fault types, fault locations, and fault conditions, together with pre- and post-fault voltage and current phasor measurements and associated metadata, providing information suitable for different fault diagnostic tasks.
\item The dataset can be used to develop and evaluate fault detection, classification, and localization approaches using classical machine-learning, deep-learning, and graph-based learning methods, as well as to compare the robustness of different approaches under changing network conditions.
\item  The data can benefit researchers working in distribution-system protection, monitoring, DER integration, and data-driven fault analytics by providing a reusable simulation resource for method development, comparative studies, and reproducible research.
\end{itemize}

\section{Background}
\label{sec:Background}
Distribution grids have become smarter networks as DERs such as PV generation, EVs, and energy storage are increasingly integrated. In standard distribution networks, electricity is mainly supplied from the upstream grid to downstream consumers, leading to more predictable power-flow patterns. With the increase in DER penetration, generation and consumption are distributed throughout the network, and the loading of the feeders, voltage profiles, and direction of power flow will be more variable~\cite{Ahsan2025}. EV charging introduces additional time-varying demand, while local generation and V2G operation can further modify the net power exchanged at individual buses. Consequently, the operating state of a distribution network can vary considerably over time and across different levels of DER integration~\cite{SECCHI2023101120}.

These changes also have important implications for fault diagnostics. Fault-induced voltage and current variations depend not only on the fault type and location, but also on the network operating condition at the time of the event. Changes in loading and DER power injection can therefore influence the electrical measurements available for detecting, classifying, and locating faults. As distribution networks become increasingly dynamic, fault diagnostic methods need to be evaluated across a sufficiently broad range of operating conditions rather than under a limited set of fixed loading and generation assumptions~\cite{Mora2015}.

Data-driven approaches based on artificial intelligence (AI), including machine-learning and deep-learning, have gained increasing attention for fault diagnostics in distribution systems because of their ability to learn complex relationships from electrical measurements~\cite{computers15020076}. These methods have been applied to fault detection, classification, and localization using voltage, current, and other system measurements. However, their development and evaluation depend strongly on the availability of sufficiently large and diverse datasets that represent different network operating and fault conditions. In particular, datasets generated under limited loading conditions or fixed DER penetration levels may not adequately capture the variability encountered in DER-integrated distribution systems.

Obtaining large and systematically labeled fault datasets from distribution system operators is challenging because actual fault events are relatively uncommon, measurement availability can be limited, and controlled generation of different fault conditions is generally impractical. Simulation-based dataset generation therefore provides a practical approach for systematically studying fault behavior across different operating conditions, fault types, locations, and DER integration levels~\cite{Kouraichi11421303}.

To support research in this area, this work presents \textbf{DER-Fault}, a simulation-based fault dataset developed using the Iowa 240-bus distribution network~\cite{data_240}. The dataset incorporates representative operating conditions selected from annual load data together with multiple PV, EV, and V2G integration scenarios. Normal and faulted operating conditions are simulated, and pre- and post-fault voltage and current phasor measurements are recorded along with associated operating-condition, DER-scenario, and fault metadata. The resulting dataset is intended to provide a reusable resource for the development, evaluation, and comparison of data-driven fault diagnostic approaches.

\section{Data Description}
\label{sec:Data_Description}
The dataset contains simulation-based electrical measurements generated from a 240-bus distribution system under varying loading, DER, and fault conditions. Sixty representative operating conditions are considered together with six DER scenarios incorporating PV generation, EV charging, and V2G operation. The released dataset contains 19,800 simulation cases representing normal operation and ten fault classes. Pre- and post-fault voltage and line-current phasor measurements are recorded at selected measurement locations within the distribution system.

A summary of the dataset specifications is provided in Table~\ref{tab:spec}. Detailed descriptions of the distribution-system model, measurement configuration, representative load selection, DER modeling, fault generation, and dataset validation are presented in Section~\ref{sec:design}.

\begin{table}[htbp]
    \centering
    \begin{tabularx}{\textwidth}{
    >{\raggedright\arraybackslash}p{0.23\textwidth}
    >{\raggedright\arraybackslash}X
}
    \toprule
       \textbf{Specification}  & \textbf{Description} \\
       \midrule
         Subject area & Electrical engineering; power and energy systems\\
         Specific subject area & Fault diagnostics in DER-integrated distribution systems\\
         Type of data & Numerical simulation data, tabular data, and metadata\\
         Data generation & OpenDSS-based simulation of normal and faulted operating conditions under varying loading and DER integration scenarios\\
          Distribution system & Distribution grid in Midwest U.S.~\cite{data_240}\\
           Data description & The dataset contains 19,800 simulation cases generated under 60 representative operating conditions and six DER scenarios incorporating different levels of PV generation, EV charging, and V2G operation. The data represent normal operation and ten fault classes and include pre- and post-fault voltage and line-current magnitude and phase-angle measurements obtained from selected measurement locations within the distribution system. The dataset contains two primary components: electrical measurement data and the corresponding sample-level metadata. Each row of the measurement dataset represents one simulation case and is linked to its metadata through a unique sample identifier. The metadata provide the information required to associate each measurement record with its loading condition, DER scenario, and fault condition.\\
            Data format & CSV\\
            Primary data source & https://wzy.ece.iastate.edu/Testsystem.html\\
            Data accessibility & Repository name: Zenodo \\
            & Data identification number: https://doi.org/10.5281/zenodo.22904220\\
            & Direct URL to data: https://zenodo.org/records/22904220 \\
            \bottomrule
    \end{tabularx}
    \caption{Data Specifications}
    \label{tab:spec}
\end{table}
\subsection{Measurement Data}
The measurement dataset is structured such that each row represents a single simulation case, and each case is identified by a unique \textit{SampleID}. The other columns are electrical measurements taken before and after the fault. The voltage measurements are arranged first, followed by the line-current measurements. For each measured bus phase, the voltage features include the pre- and post-fault voltage magnitude (Vmag\_pre and Vmag\_post) and phase angle (Vang\_pre and Vang\_post). The corresponding column names identify the bus, phase, measurement quantity, and operating state; for example, bus3003.1\_Vmag\_pre represents the pre-fault voltage magnitude of phase 1 at bus 3003.

The voltage measurements are followed by line current measurements. For each monitored line, current magnitude and phase angle are recorded at the associated terminal and phase under both pre-fault and post-fault conditions. The column names indicate the line, terminal, phase, and measurement type. For example, Line.L\_3161\_3162\_T1\_Ph1\_Imag\_pre denotes the pre-fault current magnitude for phase 1 at terminal 1 of line L\_3161\_3162. This structure allows each current measurement to be directly associated with its location in the distribution network.

\subsection{Labels and Metadata}
The measurement dataset includes a separate labels and metadata file that contains operating DER and fault information for each simulation case. A unique \textit{SampleID} links each measurement record to its corresponding metadata. The metadata includes the representative operating condition and timestamp, DER scenario parameters, and fault information. The dataset contains 11 classes, with label 0 representing normal operation and labels 1–10 corresponding to the ten simulated fault types. The complete metadata fields are summarized in Table~\ref{tab:labels}. 

\begin{table}[htbp]
    \centering
    \begin{tabularx}{\textwidth}{
    >{\raggedright\arraybackslash}p{0.23\textwidth}
    >{\raggedright\arraybackslash}X}
    \toprule
       \textbf{Fields}  & \textbf{Description} \\
       \midrule
         SampleID & Unique identifier linking the metadata to the corresponding measurement sample\\
         Representative\_Load\_Index & Index of the selected representative operating condition (1–60)\\
         Timestamp &Timestamp of the selected operating condition\\
        Hour  & Hour of day corresponding to the selected operating condition\\
          Load\_Category & Operating-condition category: peak, off-peak, transition, typical weekday, or typical weekend\\
           Scenario\_ID & DER scenario identifier\\
            PV\_Penetration & PV penetration level\\
            EV\_Penetration & EV penetration level \\
            V2G\_Participation& V2G participation level\\
            PV\_Profile\_Mult & Normalized PV generation multiplier corresponding to the selected hour\\
            EV\_Charge\_Profile\_Mult & Normalized EV charging multiplier corresponding to the selected hour\\
            V2G\_Profile\_Mult & Normalized V2G discharging multiplier corresponding to the selected hour\\
            Fault\_Type & Fault type or normal operating condition\\
            Fault\_Label & Numerical class label (0–10)\\
            Fault\_Bus & Bus associated with the simulated fault case\\
            Resistance & Fault resistance in ohms\\ 
            \bottomrule
    \end{tabularx} 
    \caption{Description of the labels and metadata fields}
    \label{tab:labels}
\end{table}

\section{Experimental Design, Materials and Methods}
\label{sec:design}
The dataset generation framework was designed to capture fault behavior under diverse network loading and DER operating conditions. The overall procedure includes distribution-system modeling, measurement configuration, representative load selection, DER scenario modeling, fault simulation, and dataset validation. The following subsections describe each stage of the dataset generation process.
\subsection{Distribution System and Simulation Setup}
\begin{figure}[htbp]
    \centering
    \includegraphics[width=1\linewidth]{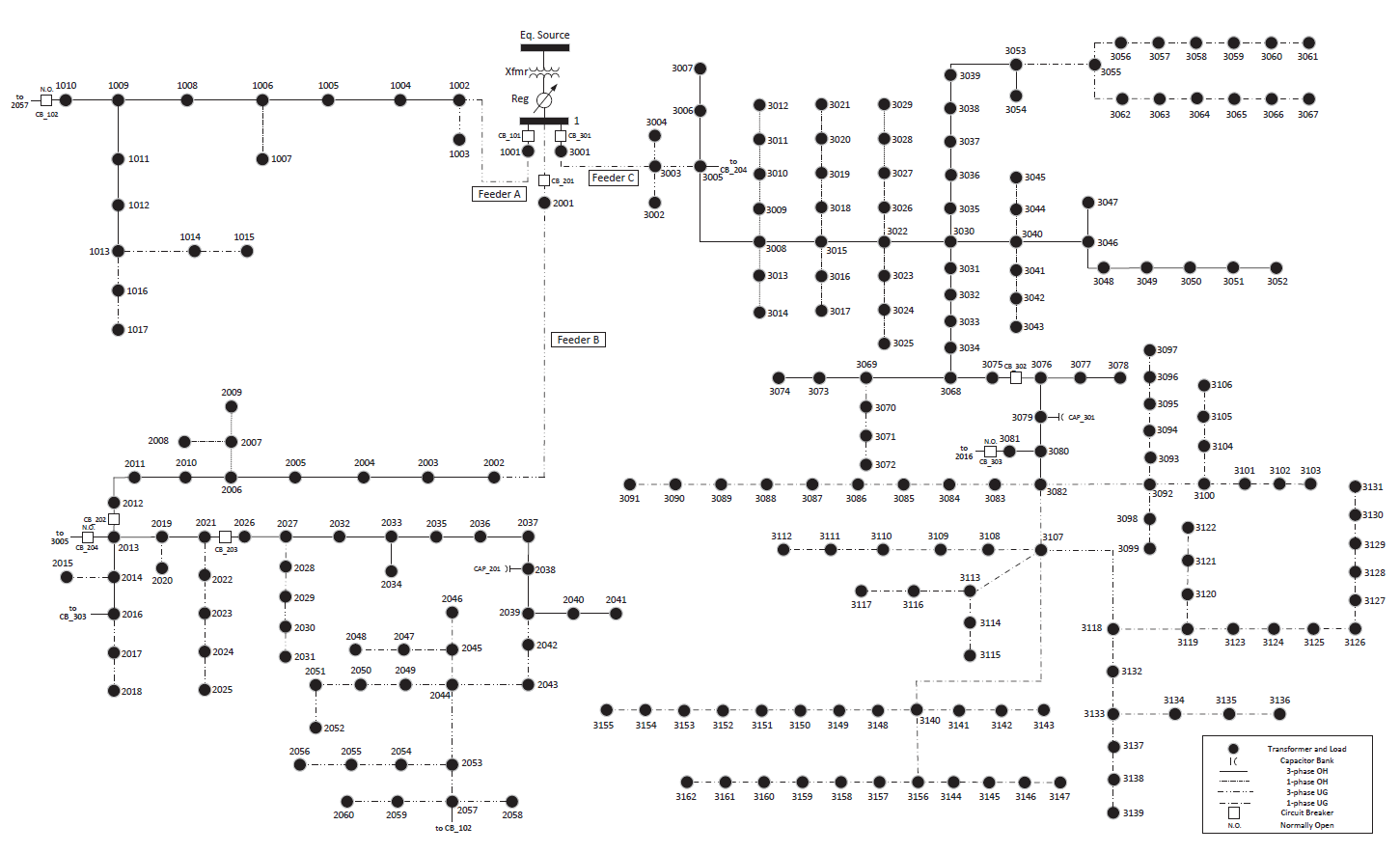}
    \caption{Single-line diagram of test system~\cite{data_240}.}
    \label{fig:cct}
\end{figure}

The Iowa 240-bus distribution system was used as the test network for dataset generation. The network model was implemented in OpenDSS and simulated using the OpenDSSDirect Python interface. The original distribution-system model and annual load data were obtained from~\cite{data_240}. The network topology is shown in Figure~\ref{fig:cct}. The system consists of a three-phase distribution network with both three-phase and single-phase connections. The load dataset provides bus-level active and reactive power demand at hourly resolution over one year, resulting in 8,760 operating conditions. These data were used to establish the baseline loading conditions of the distribution system.
For each selected operating condition, the corresponding active and reactive power demand at individual buses was applied to the OpenDSS model. The baseline loading conditions were subsequently modified according to the DER scenario under consideration. For each simulation case, the network was first solved under the specified operating condition to obtain the pre-fault electrical quantities. Following fault application, the network was solved again to obtain the corresponding post-fault quantities. Voltage and line-current phasor measurements obtained from these simulations were used to construct the dataset.
\subsection{Measurement Configuration} 
Phasor measurement units (PMUs) were used to define the measurement locations for dataset generation. A total of 93 PMUs were considered in the 240-bus distribution system, corresponding to the full-observability measurement configuration adopted in this study. Under the adopted one-hop observability criterion, a bus is considered observable if a PMU is installed either at the bus itself or at one of its directly connected neighboring buses. The same measurement configuration was used consistently throughout the dataset-generation process. The complete list of the 93 PMU locations is provided with the publicly available dataset on Zenodo~\cite{mohamed_razick_2026_2290422}. 

Prior to simulation, the selected PMU locations were verified against the buses available in the OpenDSS network model, and the phase configuration of each measurement bus was identified.
At each PMU-equipped bus, voltage magnitude and phase angle were obtained for all available phases under both pre-fault and post-fault conditions. Accordingly, the voltage measurements associated with each measured bus phase $i$ can be represented as

{\centering $v_i = \left[V_{i, \text{mag}}^{\text{pre}}, V_{i, \text{ang}}^{\text{pre}}, V_{i, \text{mag}}^{\text{post}}, V_{i, \text{ang}}^{\text{post}}\right].$\par}

Current phasor measurements were also obtained for the monitored lines associated with the selected measurement buses. For each monitored line phase, current magnitude and phase angle were recorded under both pre-fault and post-fault conditions, resulting in

{\centering$i_{l, \phi} = \left[I_{{l, \phi}, \text{mag}}^{\text{pre}}, I_{{l, \phi}, \text{ang}}^{\text{pre}}, I_{{l, \phi}, \text{mag}}^{\text{post}}, I_{{l, \phi}, \text{ang}}^{\text{post}}\right].$\par}

where $l$ and $\phi$ denote the monitored line and phase, respectively. The implemented measurement functions retrieve voltage magnitude and angle at the selected buses and current magnitude and angle for the monitored line phases. 

\subsection{Load Selection}\label{ssec:load}
The annual load profiles used in this work were obtained from~\cite{data_240} and contain hourly active and reactive power demand for one year, resulting in 8,760 operating conditions.

To capture annual load variability while limiting the number of simulation cases, 60 representative operating conditions were selected from these hourly load profiles. Five actual hourly conditions were retained from each month: peak, off-peak, transition, typical weekday, and typical weekend conditions. The peak and off-peak conditions correspond to the maximum and minimum monthly total feeder demand, respectively, while the transition condition represents the hour exhibiting the largest absolute one-hour change in feeder demand among the remaining conditions. The typical weekday and weekend conditions were selected as the actual hourly operating points with total feeder demand closest to the corresponding monthly mean weekday and weekend demand. 

The resulting set contains 60 unique operating conditions, with 12 conditions from each of the five categories. For every selected timestamp, the original bus-level active and reactive power values were retained, preserving both the spatial distribution of loading across the network and the corresponding active–reactive power pairing.

\subsection{DER Scenario Modeling}
To evaluate the dataset under different levels of DER integration, six operating scenarios were considered by varying PV penetration, EV penetration, and V2G participation. The considered scenarios are summarized in Table~\ref{tab:der}. The first scenario represents the baseline network without DER integration, while the remaining scenarios introduce different combinations of PV generation, EV charging, and V2G operation.
\begin{table}[htbp]
\centering
\begin{tabular}{lccc}
\hline
Scenario & PV Penetration (\%) & EV Penetration (\%) & V2G Participation (\%) \\
\hline
S1 & 0  & 0  & 0  \\
S2 & 30 & 0  & 0  \\
S3 & 50 & 20 & 0  \\
S4 & 70 & 30 & 0  \\
S5 & 70 & 30 & 30 \\
S6 & 30 & 50 & 50 \\
\hline\\
\end{tabular}
\caption{DER penetration scenarios considered in this study.}
\label{tab:der}
\end{table}

PV penetration was defined with respect to the peak feeder active power demand. For a specified PV penetration level $\alpha_{\mathrm{PV}}$, the total installed PV capacity was determined as
\begin{equation}
    P_{\mathrm{PV}}^{\mathrm{rated}} = \alpha_{\mathrm{PV}} P_{\mathrm{peak}},
\end{equation}
where $P_{\mathrm{peak}}$ is the maximum total feeder demand among the selected operating conditions. The total PV capacity was distributed among the load buses according to their average active-power demand, such that buses with larger average loads were assigned proportionally larger PV capacities. 
The PV output at bus $i$ and time $t$ was determined as
\begin{equation}
    P_{i, t}^{\mathrm{PV}} =P_{PV, i}^{\mathrm{rated}} f_{\mathrm{PV}}(t).
\end{equation}

EV penetration was defined as the proportion of the assumed residential population equipped with an EV. A total of 1,120 households was considered, with one EV assigned to each participating household in the distribution system. Each EV was modeled with a rated charging power of 7.2 kW and a power factor of 0.98~\cite{lillebo2019}. The total EV charging demand at time $t$ was calculated as
\begin{equation}
    P_{t}^{\mathrm{EV}} = \alpha_{\mathrm{EV}} N_{\mathrm{HH}} P_{\mathrm{ch}} f_{\mathrm{EV}}(t),
\end{equation}
where $\alpha_{\mathrm{EV}}$ is the EV penetration level and $N_{\mathrm{HH}}=1120$ is the assumed number of households $P_{\mathrm{ch}}=7.2 \mathrm{ kW}$ is the rated charging power per EV, and $f_{\mathrm{EV}}(t)$ is the normalized EV charging profile value at time $t$. The resulting EV population was distributed among the load buses in proportion to their average active-power demand.

For scenarios incorporating V2G operation, the V2G participation level represents the proportion of the EV population capable of supplying power back to the distribution network. A rated discharge power, $P_{\mathrm{dis}}$ of 5 kW per participating EV was considered. Accordingly, the total V2G power injection at time $t$ was calculated as
\begin{equation}
    P_{t}^{\mathrm{V2G}} = \alpha_{\mathrm{V2G}} N_{\mathrm{EV}} P_{\mathrm{dis}} f_{\mathrm{V2G}}(t),
\end{equation} 
where $\alpha_{\mathrm{V2G}}$ is the V2G participation level, $N_{\mathrm{EV}} =\alpha_{\mathrm{EV}} N_{\mathrm{HH}}$ is the EV population corresponding to the specified EV penetration level, and $f_{\mathrm{V2G}}(t)$ is the normalized V2G discharge profile value at time $t$.

The temporal variation in PV generation, EV charging, and V2G operation was represented using the normalized 24-hour profiles constructed based on representative temporal patterns reported in the literature~\cite{COOK20181037, Kurukuru07022023}. The corresponding hourly profile values were used to scale the PV generation, EV charging demand, and V2G power injection for each selected operating condition.

For each representative operating condition, the baseline bus-level active and reactive power demands were combined with the corresponding PV generation, EV charging demand, and V2G power injection. The resulting net active power at bus $i$ and time $t$ was calculated as
\begin{equation}
    P_{i,t}^{\mathrm{net}} = P_{i,t}^{\mathrm{load}} + P_{i,t}^{\mathrm{EV}} - P_{i,t}^{\mathrm{PV}} - P_{i,t}^{\mathrm{V2G}},
\end{equation}
where $P_{i,t}^{\mathrm{load}}$, $P_{i,t}^{\mathrm{EV}}$, $P_{i,t}^{\mathrm{PV}}$, and $P_{i,t}^{\mathrm{V2G}}$ represent the original bus demand, EV charging demand, PV generation, and V2G power injection, respectively. Reactive power was similarly determined according to the specified power factors. The resulting active and reactive power values were applied as the net bus loads in the OpenDSS model for each simulation condition.

\subsection{Fault Generation}
The dataset considers normal operation and ten short-circuit fault classes, including three single-line-to-ground faults (LGA, LGB, and LGC), three line-to-line faults (LLAB, LLBC, and LLAC), three double-line-to-ground faults (LLABG, LLBCG, and LLACG), and one three-phase-to-ground fault (LLLG)~\cite{computers15020076}. Together with the normal operating condition, these result in 11 classes. Numerical labels from 0 to 10 were assigned to the classes, with label 0 representing normal operation and labels 1–10 corresponding to LGA, LGB, LGC, LLAB, LLBC, LLAC, LLABG, LLBCG, LLACG, and LLLG, respectively.

Faults were simulated at selected buses of the distribution network using the phase connections corresponding to each fault type. Single-line-to-ground faults were applied individually to phases A, B, and C; line-to-line faults were applied between phases AB, BC, and AC; double-line-to-ground faults were applied to AB-G, BC-G, and AC-G; and the three-phase-to-ground fault involved all three phases and ground. For each faulted case, the fault resistance $R_f$ was independently sampled from a uniform distribution between 0.05 and 25 $\Omega$.

For each case, pre-fault voltage and current measurements were recorded, followed by fault application and post-fault measurement acquisition. The fault-generation procedure was repeated across the 60 representative loading conditions and six DER scenarios. With five samples generated for each of the 11 classes, the resulting dataset contains 19,800 simulation cases.

\subsection{Technical Validation of Dataset}
The generated dataset was technically validated from four complementary perspectives: coverage of the selected operating conditions, coverage of the considered DER scenarios, phase-specific characteristics of the simulated fault types, and the relationship between fault resistance and voltage-phasor disturbance severity.
\subsubsection{Operating condition coverage}
\begin{figure}[htbp]
    \centering
    \includegraphics[width=1\linewidth]{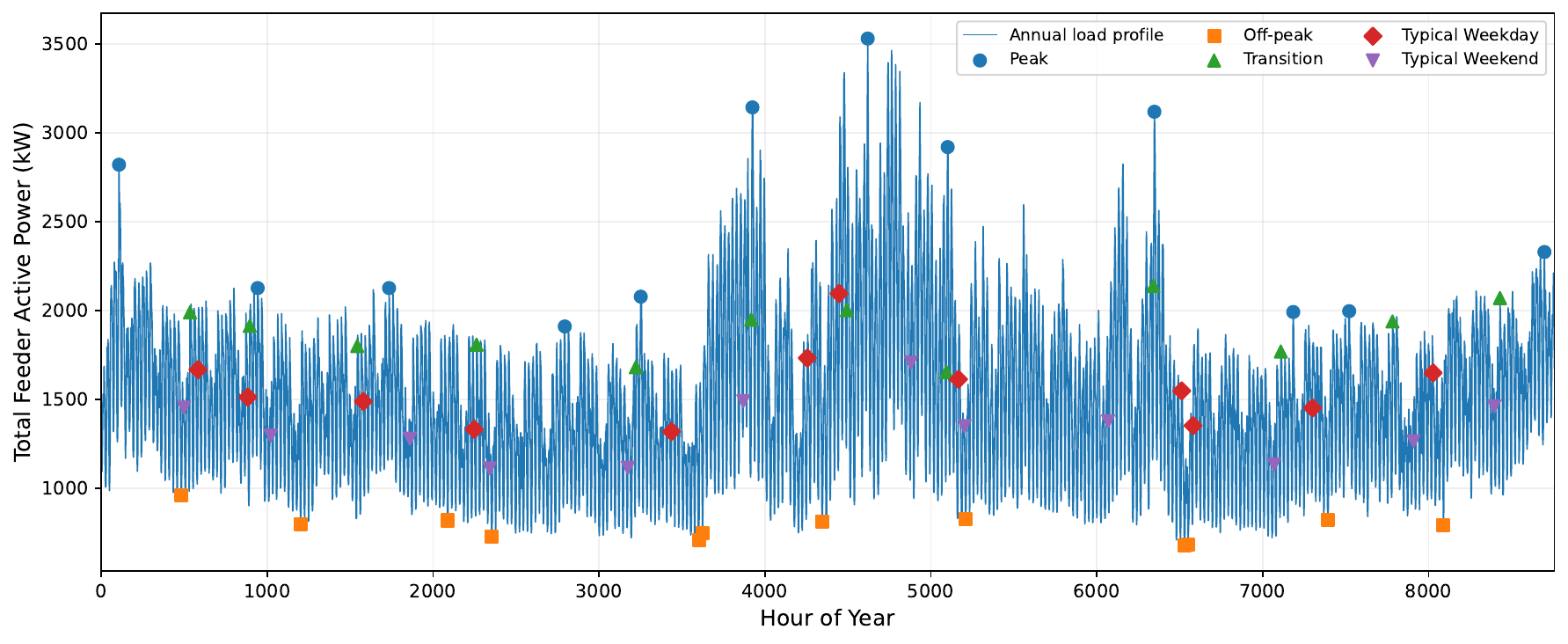}
    \caption{Annual total feeder active-power profile and the 60 representative operating conditions selected for dataset generation.}
    \label{fig:load}
\end{figure}

The coverage of the selected operating conditions was evaluated against the complete annual load profile to verify that the reduced set captures diverse loading conditions throughout the year. Figure~\ref{fig:load} shows the total feeder active power over the 8,760 hourly operating conditions together with the 60 representative conditions selected using the procedure described in Section~\ref{ssec:load}. The selected conditions are distributed throughout the annual profile and include monthly peak, off-peak, transition, typical weekday, and typical weekend operating conditions.

As shown in Figure~\ref{fig:load}, the selected conditions span both low- and high-loading periods as well as intermediate operating levels across the year. In particular, the monthly peak and off-peak conditions capture loading extremes, while the transition and typical weekday/weekend conditions provide additional coverage of rapidly changing and commonly occurring operating states. This confirms that the reduced set preserves a broad range of annual loading conditions while substantially reducing the number of operating points required for fault simulation. 

\subsubsection{DER scenario validation}
The modeled DER operating conditions were evaluated by examining the resulting PV generation, EV charging demand, and V2G power injection across the 60 representative operating conditions. Table~\ref{tab:der2} summarizes the range and median of the corresponding active power for each DER scenario. DER penetration scenarios are described in Table~\ref{tab:der}.  
\begin{table}[htbp]
\centering
\begin{tabular}{c cc cc cc}
\hline
\textbf{Scenario} &
\multicolumn{2}{c}{PV} &
\multicolumn{2}{c}{EV} &
\multicolumn{2}{c}{V2G} \\
\cline{2-3} \cline{4-5} \cline{6-7}
&
Range (kW) & Median (kW) &
Range (kW) &Median (kW) &
Range (kW) & Median (kW) \\
\hline

S1 & 0 - 0 & 0.00 & 0 - 0 & 0.00 & 0 - 0& 0.0 \\
S2 & 0 - 1,058.87 & 317.66 &0 - 0 & 0.00 & 0 - 0 & 0.0 \\
S3 & 0 - 1,764.78 & 529.43 & 564.48 - 1,612.80 & 1,193.47 & 0 - 0 & 0.0 \\
S4 & 0 - 2,470.70 & 741.21 & 846.72- 2,419.20 & 1,790.21 & 0 - 0& 0.0 \\
S5 & 0 - 2,470.70 & 741.21 & 846.72 - 2,419.20 & 1,790.21 & 0 - 454.50 & 50.52 \\
S6 & 0 - 1,058.87 & 317.66 & 1411.20 - 4,032.00 & 2,983.68 & 0 -- 1,260.00 & 140.04 \\

\hline
\end{tabular}
\caption{DER power levels across the representative operating conditions.}
\label{tab:der2}
\end{table}

The variation in power within each scenario results from the time-dependent PV, EV charging, and V2G profiles associated with the selected operating conditions. For example, PV generation ranges from zero to 2470.7 kW in S4 and S5, while EV charging demand ranges from 1411.2 to 4032.0 kW in S6. These results confirm that the generated dataset incorporates varying DER operating levels across the representative operating conditions rather than fixed DER power values.

\subsubsection{Phase-specific fault-characteristic validation}
The physical consistency of the generated fault measurements was evaluated by examining the phase-specific voltage response associated with each fault type. Since both voltage magnitude and phase angle are available in the dataset, the measurements were represented as complex voltage phasors. For each measured phase, the normalized change between the pre-fault and post-fault voltage phasors was calculated as
\begin{equation}
    D_{i, \phi} = \frac{|\mathbf{V}_{i,\phi}^{\mathrm{post}}| - |\mathbf{V}_{i,\phi}^{\mathrm{pre}}|}{|\mathbf{V}_{i,\phi}^{\mathrm{pre}}|},
\end{equation}
where $\mathbf{V}_{i,\phi}^{\mathrm{pre}}$ and $\mathbf{V}_{i,\phi}^{\mathrm{post}}$ represent the pre- and post-fault voltage phasors at bus $i$ and phase $\phi$, respectively. The combined three-phase voltage-phasor disturbance at measurement bus $i$ was then calculated as
\begin{equation}
    D_i = \sqrt{\frac{D_{i,A}^{2}+D_{i,B}^{2}+D_{i,C}^{2}} {3}}.
\end{equation}
For each fault case, the measurement bus with the maximum combined three-phase voltage disturbance, $D_{\mathrm{max}}$, was identified, and the phase-specific changes at that bus were retained. The median normalized phasor change was then calculated across all samples of each fault type.

\begin{figure}[htbp]
    \centering
    \includegraphics[width=0.5\linewidth]{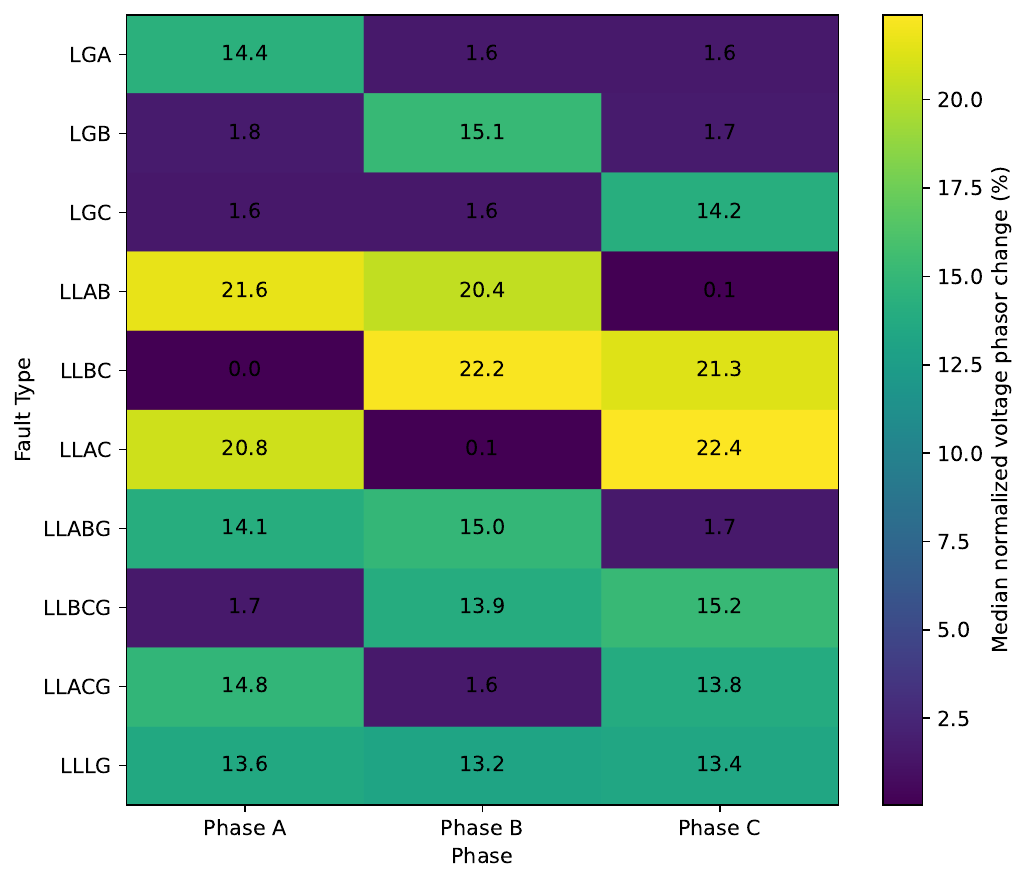}
    \caption {Normalized voltage phasor change across the three phases for each simulated fault type}
    \label{fig:heatmap}
\end{figure}

As shown in Figure~\ref{fig:heatmap}, the resulting responses exhibit clear phase-dependent characteristics consistent with the applied faults. Single-line-to-ground faults predominantly affect their corresponding faulted phase. Similarly, line-to-line and double-line-to-ground faults produce the largest phasor changes in the corresponding two faulted phases, while the remaining phase experiences a substantially smaller change. For the three-phase-to-ground fault, comparable changes are observed across all three phases. For example, the median phasor changes for the LLLG fault are 13.6\%, 13.2\%, and 13.4\% for phases A, B, and C, respectively. These results demonstrate that the generated measurements preserve the expected phase-specific characteristics of the simulated fault conditions.

\subsubsection{Fault-resistance sensitivity}
\begin{figure}[htbp]
    \centering
    \includegraphics[width=0.7\linewidth]{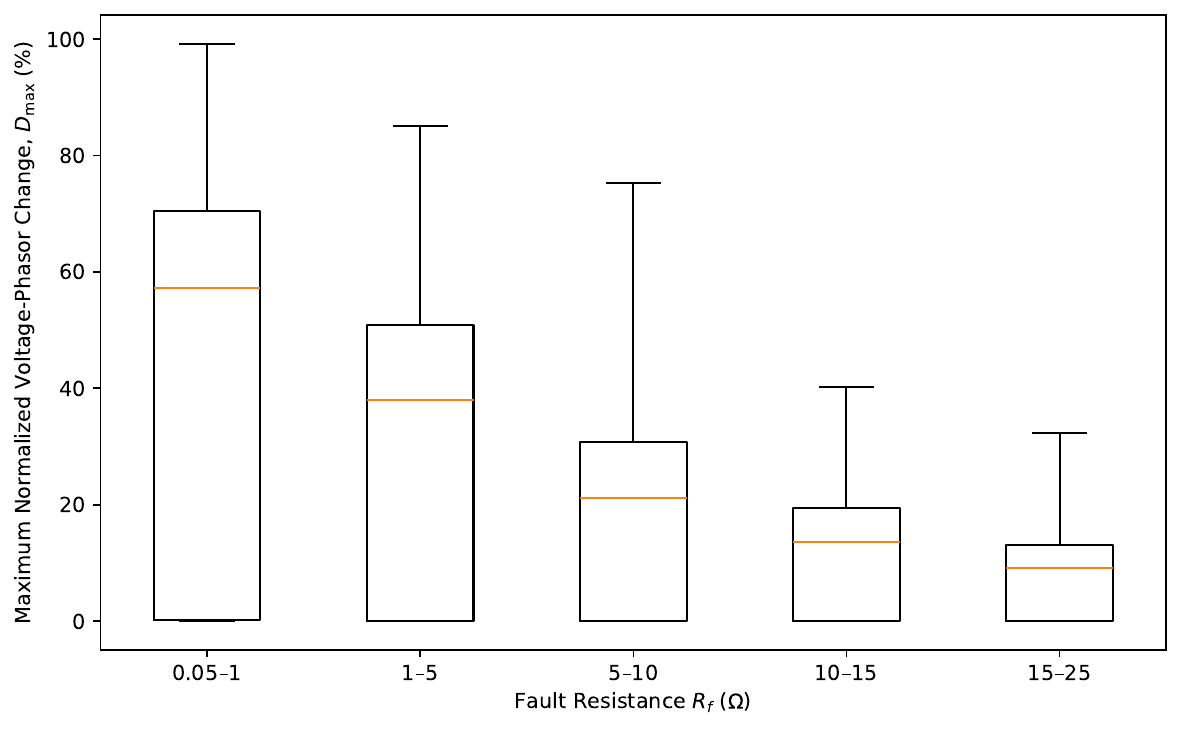}
    \caption{Distribution of $D_{\max}$ across different fault resistance ranges}
    \label{fig:resistance}
\end{figure}

To further assess the physical consistency of the simulated fault measurements, the relationship between fault resistance and voltage-phasor disturbance severity was examined. The maximum normalized voltage-phasor change, $D_{\max}$, was calculated for each fault case and grouped according to fault resistance. As illustrated in Figure~\ref{fig:resistance}, the median $D_{\max}$ decreased from 57.3\% for the 0.05–1 $\Omega$ range to 9.1\% for the 15–25 $\Omega$ range as fault resistance increased. This trend is consistent with the expected reduction in fault-induced electrical disturbance as fault resistance increases\cite{KIM09032005}. The variation within each resistance range reflects the additional influence of fault type, fault location, loading condition, and DER operating scenario.

Overall, the validation results show that the dataset captures variations in operating conditions and DER scenarios while preserving the expected phase-specific fault characteristics and fault resistance effects. These results support the technical consistency of the generated dataset across the considered operating and fault conditions.

\section{Limitations}
\label{sec:Limitations}
The dataset is generated through simulation using a single distribution system model and therefore does not capture all characteristics of real-world distribution networks. DER operation is represented using predefined PV, EV charging, and V2G profiles, while detailed inverter dynamics, battery state-of-charge behavior, and control strategies are not considered. Fault simulations are restricted to the fault types and resistance range considered in this study. In addition, the dataset contains steady-state pre- and post-fault phasor measurements rather than high-resolution transient waveforms. These limitations should be considered when applying the dataset to other network configurations or operating conditions.

\bibliographystyle{unsrt}  
\bibliography{references}

@ARTICLE{Ahsan2025,
  author={Ahsan, S M and Iqbal, M A and Hussain, A and Musilek, P},
  journal={IEEE Access}, 
  title={Adaptive Pricing-Based Optimized Resource Utilization in Networked Microgrids}, 
  year={2025},
  volume={13},
  number={},
  pages={34483-34495},
  doi={10.1109/ACCESS.2025.3543760}}

@Article{computers15020076,
AUTHOR = {M Razick, F R and Musilek, P},
TITLE = {Deep Learning for Short-Circuit Fault Diagnostics in Power Distribution Grids: A Comprehensive Review},
JOURNAL = {Computers},
VOLUME = {15},
YEAR = {2026},
NUMBER = {2},
ARTICLE-NUMBER = {76},
URL = {https://www.mdpi.com/2073-431X/15/2/76},
ISSN = {2073-431X},
DOI = {10.3390/computers15020076}
}

@article{Mora2015,
author = {Mora-Flórez, Juan J. and Herrera-Orozco, Ricardo A. and Bedoya-Cadena, Andres F.},
title = {Fault location considering load uncertainty and distributed generation in power distribution systems},
journal = {IET Generation, Transmission \& Distribution},
volume = {9},
number = {3},
pages = {287-295},
doi = {https://doi.org/10.1049/iet-gtd.2014.0325},
url = {https://ietresearch.onlinelibrary.wiley.com/doi/abs/10.1049/iet-gtd.2014.0325},
year = {2015}
}

@article{SECCHI2023101120,
title = {Smart electric vehicles charging with centralised vehicle-to-grid capability for net-load variance minimisation under increasing EV and PV penetration levels},
journal = {Sustainable Energy, Grids and Networks},
volume = {35},
pages = {101120},
year = {2023},
issn = {2352-4677},
doi = {https://doi.org/10.1016/j.segan.2023.101120},
url = {https://www.sciencedirect.com/science/article/pii/S2352467723001285},
author = {M. Secchi and G. Barchi and D. Macii and D. Petri}
}

@ARTICLE{Kouraichi11421303,
  author={Kouraichi, Maher and Mansouri, Majdi and Sakly, Anis and Trabelsi, Mohamed and Mnif, Faiçal},
  journal={IEEE Access}, 
  title={A Comprehensive Dataset and Simulation Framework for Fault Analysis in Power Transmission Systems: Dataset Description and Statistical Insights}, 
  year={2026},
  volume={14},
  number={},
  pages={37968-38008},
  doi={10.1109/ACCESS.2026.3670392}}

@INPROCEEDINGS{data_240,
  author={Bu, Fankun and Yuan, Yuxuan and Wang, Zhaoyu and Dehghanpour, Kaveh and Kimber, Anne},
  booktitle={2019 North American Power Symposium (NAPS)}, 
  title={A Time-Series Distribution Test System Based on Real Utility Data}, 
  year={2019},
  volume={},
  number={},
  pages={1-6},
  doi={10.1109/NAPS46351.2019.8999982}}

@article{COOK20181037,
title = {Modeling constraints to distributed generation solar photovoltaic capacity installation in the US Midwest},
journal = {Applied Energy},
volume = {210},
pages = {1037-1050},
year = {2018},
issn = {0306-2619},
doi = {https://doi.org/10.1016/j.apenergy.2017.08.108},
url = {https://www.sciencedirect.com/science/article/pii/S0306261917311297},
author = {Tyson Cook and Lee Shaver and Paul Arbaje}}

@article{lillebo2019,
author = {Lillebo, Martin and Zaferanlouei, Salman and Zecchino, Antonio and Farahmand, Hossein},
title = {Impact of large-scale EV integration and fast chargers in a Norwegian LV grid},
journal = {The Journal of Engineering},
volume = {2019},
number = {18},
pages = {5104-5108},
doi = {https://doi.org/10.1049/joe.2018.9318},
url = {https://ietresearch.onlinelibrary.wiley.com/doi/abs/10.1049/joe.2018.9318},
eprint = {https://ietresearch.onlinelibrary.wiley.com/doi/pdf/10.1049/joe.2018.9318},
year = {2019}
}

@Article{Kurukuru07022023,
  author    = {Varaha Satya Bharath Kurukuru and Mohammed Ali Khan and Rupam Singh},
  journal   = {Electric Power Components and Systems},
  title     = {Electric Vehicle Charging/Discharging Models for Estimation of Load Profile in Grid Environments},
  year      = {2023},
  number    = {3},
  pages     = {279--295},
  volume    = {51},
  doi       = {10.1080/15325008.2022.2146811},
  eprint    = {https://doi.org/10.1080/15325008.2022.2146811},
  publisher = {Taylor \& Francis},
  url       = {https://doi.org/10.1080/15325008.2022.2146811},
}

@Article{KIM09032005,
  author    = {S. D. Kim and M. M. Morcos and J. C. Gomez},
  journal   = {Electric Power Components and Systems},
  title     = {Voltage-Sag Magnitude and Phase Jump due to Short Circuits in Distribution Systems with Variable Fault Resistance},
  year      = {2005},
  number    = {5},
  pages     = {493--512},
  volume    = {33},
  doi       = {10.1080/15325000590504984},
  eprint    = {https://doi.org/10.1080/15325000590504984},
  publisher = {Taylor \& Francis},
  url       = {https://doi.org/10.1080/15325000590504984},
}

@misc{mohamed_razick_2026_2290422,
  author    = {Razick, F R M and Musilek, Petr},
  title     = {DER-Fault: A Dataset for Fault Analysis in DER- Integrated Distribution Systems},
  doi={10.5281/zenodo.22904220},
  url ={https://doi.org/10.5281/zenodo.22904220},
  version   = {1.0},
  month     = sep,
  howpublished  = {Zenodo, 10.5281/zenodo.22904220},
  year      = {2026},
}

\end{document}